\documentclass[twocolumn,showpacs,superscriptaddress,preprintnumbers,amsmath,amssymb]{revtex4}

\usepackage{graphicx}
\usepackage{dcolumn}
\usepackage{bm}
\usepackage{color}

\begin{document}

\preprint{}

\title{Weakly coupled $s = 1/2$ quantum spin singlets in Ba$_{3}$Cr$_{2}$O$_{8}$
}

\author{M. Kofu}
\author{J.-H. Kim}
\affiliation{Department of Physics, University of Virginia, Charlottesville, Virginia 22904-4714, USA}
\author{S. Ji}
\affiliation{Department of Physics, University of Virginia, Charlottesville, Virginia 22904-4714, USA}
\affiliation{NIST Center for Neutron Research, Gaithersburg, Maryland 20899-6100, USA}
\author{S.-H. Lee}
\affiliation{Department of Physics, University of Virginia, Charlottesville, Virginia 22904-4714, USA}
\author{H. Ueda}
\affiliation{Institute for Solid State Physics, University of Tokyo, Kashiwa 277-8581, Japan}
\author{Y. Qiu}
\author{H.-J. Kang}
\author{M. Green}
\affiliation{NIST Center for Neutron Research, Gaithersburg, Maryland 20899-6100, USA}
\affiliation{Department of Materials Science and Engineering, University of Maryland, College Park, Maryland 20742-6393, USA}
\author{Y. Ueda}
\affiliation{Institute for Solid State Physics, University of Tokyo, Kashiwa 277-8581, Japan}

\date{\today}

\begin{abstract}
Using single crystal inelastic neutron scattering with and without application of an external magnetic field and powder neutron diffraction, we have characterized magnetic interactions in Ba$_3$Cr$_2$O$_8$. Even without field, we found that there exist three singlet-to-triplet excitation modes in $(h,h,l)$ scattering plane. Our complete analysis shows that the three modes are due to spatially anisotropic interdimer interactions that are induced by local distortions of the tetrahedron of oxygens surrounding the Jahn-Teller active Cr$^{5+} (3d^1)$. The strong intradimer coupling of $J_0 =  2.38(2)$ meV and weak interdimer interactions ($|J_{\rm inter}| \leq 0.52(2)$~meV) makes Ba$_3$Cr$_2$O$_8$ a good model system for weakly-coupled $s = 1/2$ quantum spin dimers.
\end{abstract}

\maketitle

For the past two decades, the field of quantum magnets has provided a number of new exotic collective phenomena~\cite{Sachdev08}. Progress has always been moderated by the availability of suitable spin $1/2$ materials with distinct structural motifs. The ubiquitous Cu$^{2+} (3d^9, s = 1/2)$ ion is often relied upon as the platform for new materials design. Magnetic field-induced quantum phase transitions are a typical example. The field-induced condensation of magnons have been experimentally observed in coupled quantum (s = 1/2) dimer systems based on Cu$^{2+}$ ions, such as TlCuCl$_3$~\cite{Nikuni00,Tanaka01,Ruegg03} and BaCuSi$_2$O$_6$~\cite{Sebastian06}, which are adequately described by the Bose-Einstein condensation~(BEC) theory~\cite{Nikuni00, Giamarchi99,Matsumoto02}.
However, the robustness of such descriptions can only be truly evaluated with investigation into complementary materials. This calls for finding other good model systems of quantum (s = 1/2) dimers, especially based on non-Cu$^{2+}$ ions.

\begin{figure}[b]
\begin{center}
\includegraphics[width=0.9\hsize]{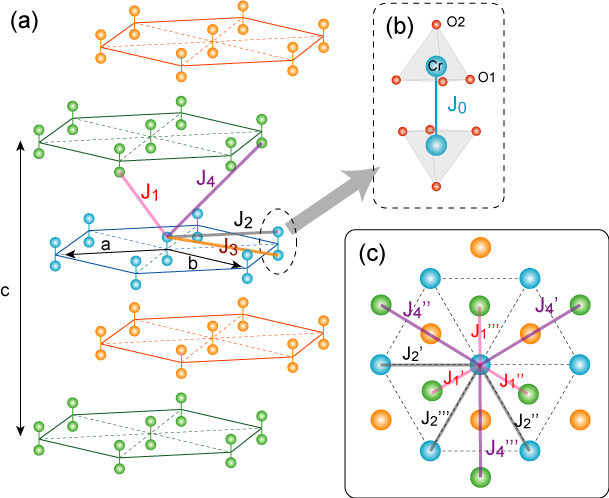}
\caption{
Schematic diagram of the crystal structure of Ba$_3$Cr$_2$O$_8$ that shows (a) the triangular network of Cr$^{5+}$ dimers in the $ab$-plane and their relative stacking along the $c$-axis in an ABCABC arrangement. (b) Two oriented Cr$^{5+}$ tetrahedra bound with four oxygen (O$^{2-}$) ions to afford the Cr$^{5+}$ dimer arrangement with magnetic exchange, $J_0$, which is at a distance of $d_0$ = 3.97\AA. (c) The $ab$-plane projection of the Cr$^{5+}$ ions, showing the alternative exchange pathways. 
}
\label{fig:crystal_structure}
\end{center}
\end{figure}

Recently, a new class of spin dimer systems has been reported with the general formula, A$_3$B$_2$O$_8$~\cite{Uchida01,Tsujii05,Nakajima07,Singh07}, where A is an alkaline earth metal such as Ba$^{2+}$ or Sr$^{2+}$ and B is a 5+ transition metal ion that forms dimers along the $c$-axis (see Fig.~\ref{fig:crystal_structure}(b)). Interestingly, a frustrating triangular lattice is formed by the dimers in the $ab$-plane (see Fig.~\ref{fig:crystal_structure}(a))~\cite{Uchida01}. For Ba$_3$Mn$_2$O$_8$ with orbitally nondegenerate Mn$^{5+} (3d^2, s = 1)$, specific heat measurements under $H$ provide experimental evidence of a condensation of magnons~\cite{Tsujii05}. In the field-induced antiferromagnetic (AFM) state, a magnetization plateau phase was observed as a function of $H$~\cite{Uchida01},  typical for frustrated magnets~\cite{Matsuda07}. A recent inelastic neutron scattering study showed that the magnetic interactions between dimers in Ba$_3$Mn$_2$O$_8$ are effectively two-dimensional in nature as a result of this interplane frustration~\cite{Stone08}. More recently, the related chromate (Ba,Sr)$_3$Cr$_2$O$_8$ with the quantum ($s = 1/2$) spin from the orbitally degenerate Cr$^{5+} (3d^1; s = 1/2)$ ion have been made~\cite{Nakajima07,Singh07}. These compositions do not exhibit long range order above 1.5~K and their bulk susceptibility data at low temperatures observes the quantum dimer model. The application of an external magnetic field at 1.6~K rapidly increases the bulk magnetization in Ba$_3$Cr$_2$O$_8$ above $H_{c1}\simeq 12$~T to its saturation value at $H_{c1}\approx  23$~T~\cite{Nakajima07}. In contrast to the manganate, no magnetization plateau was observed.

Here we report our inelastic neutron scattering data obtained from single crystals of Ba$_3$Cr$_2$O$_8$ with and without application of an external magnetic field. We have found that even without field there are three singlet-to-triplet excitation modes in $(h,h,l)$ directions, which contrasts with the single mode observed in Ba$_3$Mn$_2$O$_8$. Our detailed analysis shows that the three modes come from spatially anisotropic interdimer interactions in three different crystal domains. We have also performed neutron diffraction measurements on a powder sample and found experimental evidence of local lattice distortion that is due to the orbital degeneracy of the Cr$^{5+}$ $(3d^1)$ ion. We consider that the local distortion is the origin of the spatially anisotropic interdimer interactions. Our complete analysis of dispersions of the singlet-to-triplet excitations show that this system is an excellent model system for three-dimensional weakly coupled dimers with dominant intradimer coupling ($J_0 =  2.38(2)$~meV) and weak interdimer couplings less than 0.52(2)~meV. Thus, this system provides a new testing ground for theories of the condensation of magnons.

Single crystals of Ba$_3$Cr$_2$O$_8$ were grown by the floating zone method using a four mirror image furnace equipped with halogen lapms. Since Ba$_3$Cr$_2$O$_8$ is congruently melting, sintered stoichiometric rods were used. We have performed inelastic neutron scattering measurements on single crystals with a total weight of 450~mg at the cold-neutron triple-axis spectrometer (SPINS) and neutron diffraction measurements on a 15~g powder sample of Ba$_3$Cr$_2$O$_8$ at the BT1 neutron powder diffractometer, located at the National Institute of Standards and Technology (NIST) Center for Neutron Research.  At SPINS, the crystals were mounted in the $(hhl)$ scattering plane. A horizontally focusing analyzer mode was set at the final energy of $E_f = 5$~meV and to cover the scattering angle of about 5 degrees at a time. This increased the sensitivity of the instrument and allowed us to map out magnon dispersions along high symmetry directions. A liquid nitrogen cooled Be filter was placed before the analyzer to eliminate high order neutron contaminations. At SPINS, the samples were cooled in a liquid $^4$He cryostat for the measurement without field, and in a 11.5~T superconducting magnet for the field measurements. At BT1, a 4~K closed-cycle refrigerator (displex) was used.

\begin{figure}[htb]
\begin{center}
\includegraphics[width=0.9\hsize]{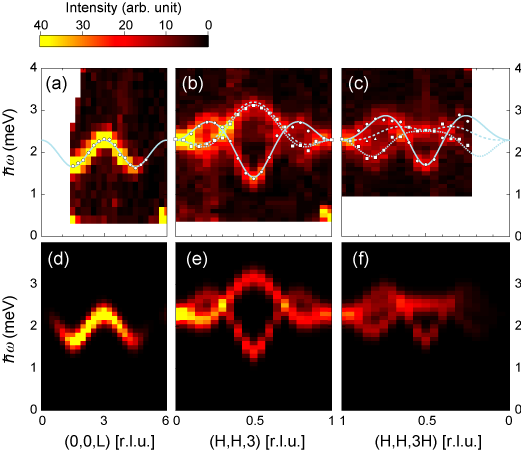}
\caption{
Contour maps of dispersion relations along (a)~$(0,0,l)$, (b)~$(h,h,3)$ and (c)~$(h,h,3h)$ directions. The data were taken by performing constant-$\bm Q$ scans of Ba$_3$Cr$_2$O$_8$ single crystals at 1.7~K. The point symbols represent the peak positions obtained by fitting the data with simple Gaussians. (d)-(e) Calculated dispersions and intensities based on a model of spatially anisotropic interdimer interactions, which is described in the text.
}
\label{fig:dispersion}
\end{center}
\end{figure}

In order to investigate the intra- and inter-dimer interactions in Ba$_3$Cr$_2$O$_8$, we have performed inelastic neutron scattering measurements on single crystals at 1.7~K. Several constant $\bm Q$-scans were performed to map the dispersion of the singlet-to-triplet excitations. Figure~\ref{fig:dispersion} shows the results along three high symmetry directions. Along the $(0,0,l)$ direction, a nearly flat excitation mode centered $\sim 2$~meV is observed with a bandwdith of $\sim 0.7$~meV (see Fig.~\ref{fig:dispersion}(a)). This dispersion is due to the weak interdimer interactions, $J_1$ and $J_4$, between the dimer planes (see Fig.~\ref{fig:crystal_structure}). The $(h,h,3)$ and $(h,h,3h)$ directions, that probe the inplane interdimer interactions, $J_2$, $J_3$, as well as the out-of-plane ones show three unexpected excitation modes (see Fig.~\ref{fig:dispersion}(b) and (c)). In contrast, Ba$_3$Mn$_2$O$_8$ displays only a single triplet excitations along all the symmetric directions measured~\cite{Stone08}. Magnetic excitations from weakly coupled dimer systems can be well explained within the random phase approximation~(RPA) that gives the dispersion relation~\cite{Stone08,Leuenberger84,Cavadini99}, $\hbar\omega (\bm Q)  \simeq  \sqrt{J_0^2+J_0~\gamma(\bm Q)}$, where $\gamma(\bm Q) =  \sum_{i} J(\bm R_i) e^{-i \bm Q \cdot  \bm R_i}$ and  $J(\bm R_i)$ represents interdimer interactions. If the interdimer interactions are spatially isotropic, as shown in Fig.~\ref{fig:crystal_structure}(a), then the four nearest interactions, $J_1$, $J_2$, $J_3$, $J_4$, gives 
\begin{eqnarray}
\gamma(\bm Q)
&=& 2J_1 [ \cos \left\{ \frac{2}{3}\pi (2h+k+l) \right\} + \cos \left\{ \frac{2}{3}\pi (-h+k+l) \right\}   \nonumber \\
&+&\cos \left\{ \frac{2}{3}\pi (-h-2k+l) \right\}] \nonumber \\
&+& 2(J_2-J3) \left[\cos\left\{2\pi h \right\}+\cos\left\{2\pi k \right\}+\cos \left\{2\pi (h+k) \right\} \right] \nonumber \\
&+& 2J_4 [ \cos \left\{ \frac{2}{3}\pi (2h+4k+l) \right\} + \cos \left\{ \frac{2}{3}\pi (2h-2k+l) \right\} \nonumber \\
&+& \cos \left\{ \frac{2}{3}\pi (-4h-2k+l) \right\}] \nonumber
\end{eqnarray}
such that without field only a single triplet excitation along any of $(h,h,l)$ directions is expected, as recently observed in the related $s = 1$ system Ba$_3$Mn$_2$O$_8$~\cite{Stone08}.

\begin{figure}[tb]
\begin{center}
\includegraphics[width=\hsize]{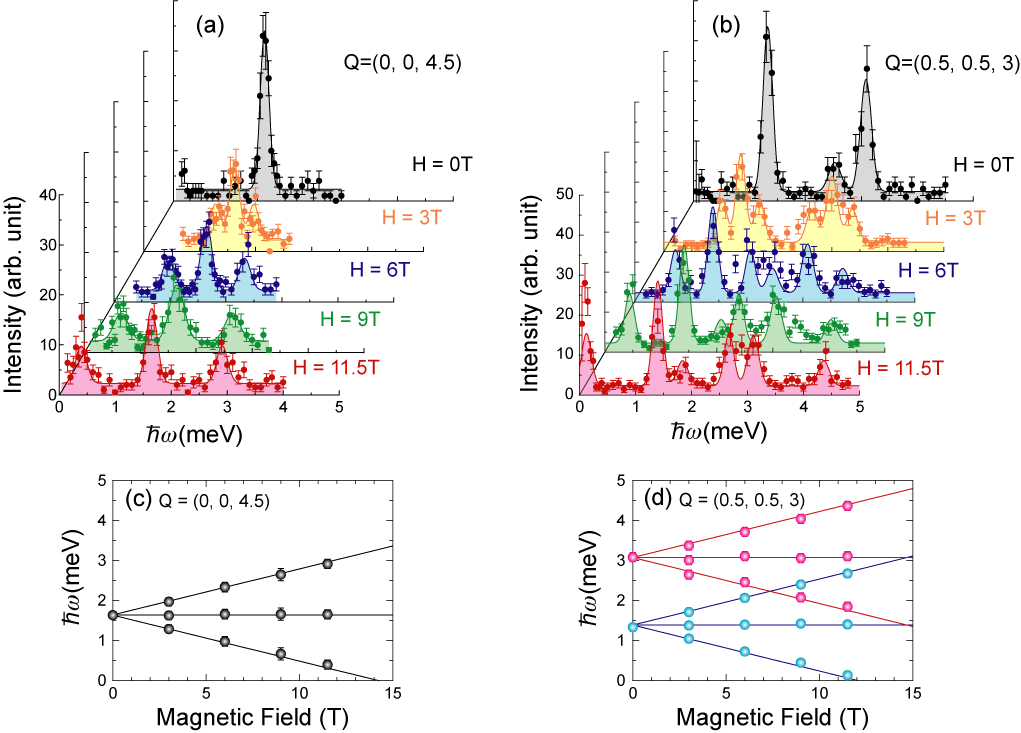}
\caption{
Constant $\bm Q$ scan with several different magnetic fields ($H$) (a) at $\bm Q = (0,0,4.5)$ and (b) at $\bm Q = (0.5,0.5,3)$ scan, and (c) $H$-dependence of the peak positions in $\hbar\omega$ (c) at $\bm Q = (0,0,4.5)$ and (d) at $\bm Q = (0.5,0.5,3)$. The lines represent the Zeeman splitting. 
}
\label{fig:Zeeman}
\end{center}
\end{figure}

This poses the question as to the origin of the complex excitations observed in this present study on Ba$_3$Cr$_2$O$_8$. One possibility would be that the triply-degenerate singlet-to-triplet excitation mode is already split into three modes in the $(h,h,l)$ scans as a result of single anisotropy terms. If this is the case then each excitation in the $(h,h,l)$ scan are not degenerate and therefore remain single under an application of an external magnetic field. This conjecture was tested by performing further inelastic neutron experiments at constant-$\bm Q$ under an external magnetic field, which are shown in Fig.~\ref{fig:Zeeman}. In Fig.~\ref{fig:Zeeman}(a), the single mode at $\bm Q = (0,0,4.5)$ at $H = 0$~T splits into three modes for finite values of field. As $H$ increases, the Zeeman energy splitting increases as expected. Furthermore, Fig.~\ref{fig:Zeeman}(b) shows the two strong modes at $\bm Q = (0.5,0.5,3)$ that appeared at $\hbar\omega=$1.4~meV and 3.05~meV without field also revert to three modes when $H$ is applied. This demonstrates unambiguously that both modes are triply degenerate singlet-to-triplet excitations.

Another possibility is that in analogy to the acoustic and optical vibrational modes that exist in a lattice with more than one ion in a unit cell, the origin might be an acoustic and two optical magnon modes as a result of the three bi-layers in the chemical unit cell. Such an optical magnon mode has been observed in Cs$_3$Cr$_2$Br$_9$, where Cr$^{3+} (3d^3; s = 3/2)$ ions form two bi-layer triangular planes~\cite{Leuenberger84}. The optical mode, however, appeared along certain symmetry directions with much weaker intensity than the acoustic mode. For Cs$_3$Cr$_2$Br$_9$, Leuenberger {\it et al.} have calculated, using a Green's function technique, the dispersions and intensities of the acoustic and optical modes. They showed that the intensity of the acoustic mode is proportional to $1+\cos(\bm\rho \cdot \bm\tau +\phi)$ while that of the optical mode is related to $1-\cos(\bm\rho \cdot \bm\tau + \phi)$ where $\bm\rho$ is the displacement vector between two different sublattices (planes), $\bm\tau$ a reciprocal lattice vector, and $\phi$ is the phase of the Fourier sum of the interplane interactions which is complex in the case of Cs$_3$Cr$_2$Br$_9$~\cite{Leuenberger84}. Their calculations were consistent with their data. We have implemented their model for the present case, Ba$_3$Cr$_2$O$_8$, which suggests the presence of one acoustic and two optical modes. However, for this case $\phi$ turns out to be zero because of the inversion symmetry at the center of the dimer and therefore the Fourier sum of the interplane interactions becomes real. Since $\bm\tau (h,k,l)$ must satisfy the reflection condition $-h+k+l = 3n$ with an integer $n$ for the $R\bar 3m$ symmetry and $\bm\rho=\left(\frac{2}{3},\frac{1}{3},\frac{1}{3}\right)$, $\cos(\bm\rho \cdot \bm\tau)$ is always 1. Therefore, the optical modes in Ba$_3$Cr$_2$O$_8$ are not present, and the observed excitations are all acoustic in nature, which is consistent with the observation of only one single triplet mode in the related compound Ba$_3$Mn$_2$O$_8$~\cite{Stone08}.

A third possibility is that the interdimer interactions are not spatially isotropic and the three modes arise from three different crystal domains. To test this scenario, we have considered a set of spatially anisotropic interdimer interactions, $J_1^{'}, J_1^{''}, J_1^{'''}, J_2^{'}, J_2^{''}, J_2^{'''}, J_4^{'}, J_4^{''}, J_4^{'''}$, as depicted in Fig.~\ref{fig:crystal_structure}(c). $J_3$ is neglected because we cannot observe $J_3$ separately but we only observe $J_2-J_3$. Then, for one crystal domain  
\begin{eqnarray}
\gamma(h,h,l)
&=& 2J_1^{'} \cos \left\{ \frac{2}{3}\pi (3h+l) \right\} + 2J_1^{''} \cos \left\{ \frac{2}{3}\pi l \right\} \nonumber \\
&+& 2J_1^{'''}\cos \left\{ \frac{2}{3}\pi (-3h+l) \right\} \nonumber \\
&+& 2(2J_2^{''}+2J_2^{''}) \cos\left\{2\pi h \right\} + 2J_2^{'''}\cos \left\{4\pi h \right\}  \nonumber \\
&+& 2J_4^{'} \cos \left\{ \frac{2}{3}\pi (-6h+l) \right\} + 2J_4^{''}\cos \left\{ \frac{2}{3}\pi l \right\} \nonumber \\
&+&2J_4^{'''}\cos \left\{ \frac{2}{3}\pi (6h+l) \right\} \nonumber
\end{eqnarray}
For the other two domains, the coupling constants are permuted: $\left\{J_{1,2,4}^{'}, J_{1,2,4}^{''}, J_{1,2,4}^{'''} \right\} \rightarrow \left\{J_{1,2,4}^{''}, J_{1,2,4}^{'''}, J_{1,2,4}^{'} \right\} \rightarrow \left\{J_{1,2,4}^{'''}, J_{1,2,4}^{'}, J_{1,2,4}^{''} \right\}$. This suggests that three domains will produce the same dispersion along the $(0,0,l)$ direction, whereas along the $(h,h,3)$ and $(h,h,3h)$ directions each domain will yield a unique dispersion as shown in different lines in Fig.~\ref{fig:dispersion}(b) and (c). Fig.~\ref{fig:dispersion}(d)-(f) are calculated dispersion relations and demonstrate that the experimental data can be simulated by this model when $J_0=2.38(1)$,  $J_1^{'}=0.08(2)$,  $J_1^{''}=-0.15(2)$,  $J_1^{'''}=0.10(1)$,  $J_2^{'}=0.10(2)$,  $J_2^{''}=0.07(2)$,  $J_2^{'''}=-0.52(2)$,  $J_4^{'}=~0.04(2)$,  $J_4^{''}=0.10(2)$,  $J_4^{''}=0.09(2)$. All values are given in meV. 

\begin{figure}[tb]
\begin{center}
\includegraphics[width=\hsize]{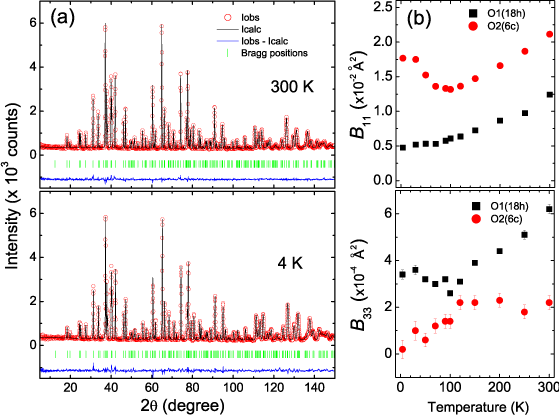}
\caption{
(a) Observed, calculated and difference intensities as a function of two-theta as obtained from Rietveld refinement of the powder neutron diffraction data at 4~K and 300~K. Expected reflections are represented by green tickmarks. (b) $T$-dependence of the anisotropic displacement parameter, $B_{ij}$, for the apical (red circles) and for the planar (black circles) oxygens.
}
\label{fig:tdep_diff}
\end{center}
\end{figure}

\begin{table}[tb]
\caption{The refined crystal structural parameters of Ba$_3$Cr$_2$O$_8$ derived from the data shown in Fig.~\ref{fig:tdep_diff}(a) using the program, Fullprof. The symmetry was found to be  $R\bar{3}m$ at all temperatures. $W$ represents the crystallographic Wyckoff position, $\chi^2 = \displaystyle\sum_i \left\{I_{{\rm obs}, i} - I_{{\rm cal}, i}\right\}^2 / \Delta I^2_{{\rm obs},i}$ and $R_F = 100\times \displaystyle\sum_{h,k,l} \left|I_{\rm obs}(h,k,l) - I_{\rm cal}(h,k,l)\right|/\displaystyle\sum_{h,k,l} \left|I_{\rm obs}(h,k,l)\right|$
where $I_{\rm obs}$ and $I_{\rm cal}$ are the observed and calculated intensity, respectively.}
\begin{ruledtabular}
\begin{tabular}{lllll}
Atom($W$) & $x$ & $y$ & $z$ & $B_{\rm{iso}}$(\AA$^2$) \\ \hline 
\multicolumn{5}{l}{300 K, $\chi^2 = 1.30 $, R$_{F^2} = 2.99$} \\
\multicolumn{5}{l}{$a=b=5.74171(6)$\AA, $c=21.38736(30)$\AA} \\
Ba1($3a$) & 0 & 0 & 0 &  1.015(36) \\
Ba2($6c$) & 0 & 0 & 0.20587(8) & 0.698(26) \\
Cr($6c$)  & 0 & 0 & 0.40704(8) & 0.689(37) \\
O1($18h$)  & 0.82832(10) & 0.17168(10) & 0.89874(3) & - \\
O2($6c$)  & 0 & 0 & 0.32847(6) & - \\ \hline
\multicolumn{5}{l}{4 K, $\chi^2 = 2.86 $, R$_{F^2} = 4.49$} \\
\multicolumn{5}{l}{$a=b=5.72077(8)$\AA, $c=21.35857(37)$\AA} \\
Ba1($3a$) & 0 & 0 & 0 & 0.466(45) \\
Ba2($6c$) & 0 & 0 & 0.20576(11) & 0.216(33) \\
Cr($6c$)  & 0 & 0 & 0.40648(11) & 0.110(47) \\
O1($18h$)  & 0.82831(13) & 0.17169(13) & 0.89866(4) & - \\
O2($6c$)  & 0 & 0 & 0.32842(8) & -  
\end{tabular} \\
Anisotropic displacement parameter ($B_{ij} \times 10^5$ \AA $^{2}$)\\
\begin{tabular}{lrrrrrr}
Atom($W$) & $B_{11}$ & $B_{22}$ & $B_{33}$ & $B_{12}$ & $B_{13}$ & $B_{23}$\\ \hline
\multicolumn{7}{c}{300 K} \\
O1(18$h$) & 1242(28) & 1242(28) & 62(2) & 945(31)  & 12(1) & -12(1) \\
O2(6$c$)  & 2113(32) & 2113(32) & 22(3) &-1056(16) &  0    & 0   \\ \hline

\multicolumn{7}{c}{4 K} \\
O1(18$h$) & 475(36) & 475(36) & 31(2) & 495(41)  & 1(5) & -1(5) \\
O2(6$c$)  & 1767(42) & 1767(42) & 12(4) &-884(21) &  0    & 0    

\end{tabular}
\end{ruledtabular}
\end{table}

A central question arises as to what is the origin of the spatially anisotropic interdimer exchange interactions in Ba$_3$Cr$_2$O$_8$? In order to address this issue, we have performed neutron diffraction measurements on a powder sample at various temperatures from 4~K to 300~K. We found that there is no evidence of a reduction of the ambient $R\bar 3m$ crystal structure symmetry on cooling down to 4~K. The displacement parameters, $B_{ij}$, of the apical oxygen, however, show anomalous behavior as a function of $T$. Firstly, a significant improvement in the Rietveld fit to the diffraction pattern is obtained with the use of anisotropic values for the displacement parameters, $B_{ij}$, the temperature dependence of which are shown in Fig.~\ref{fig:tdep_diff}(b). Secondly, upon cooling the ab-components of the displacement parameter, $B_{11} = B_{22}$, for the apical ($6c$) oxygen decreases down to $\sim 100$~K but then increases on further cooling (see red circles in Fig.~\ref{fig:tdep_diff}(b)) while the $c$-component, $B_{33}$, decreases below 100~K. In contrast, $B_{ij}$ for the planar ($18h$) oxygen shows a very weak $T$-dependence below 100~K. This suggests that the apical oxygen that is located at a high symmetric $6c (0,0,z)$ site locally moves in the ab-plane for $T < 100$~K. The local disorder is presumably a result of the $e_g$ orbital degeneracy within the Cr$^{5+} (3d^1)$ tetrahedral ion~\cite{Koo06}, which has a tendency to affect the local oxygen environment and thereby removing the degeneracy. Even though the Jahn-Teller distortion does not seem to lower the bulk crystal symmetry in Ba$_3$Cr$_2$O$_8$, the anomaly in $B_{ij}$ indicates the presence of local disorder, which induces the spatial exchange anisotropy that was observed in our inelastic neutron scattering measurements. Very recently, Chapon {\it et al.}~\cite{Chapon08} have studied a related material Sr$_3$Cr$_2$O$_8$ using a powder neutron diffraction technique and found that at 1.6~K the crystal structure is distorted to a monoclinic $C2/c$ symmetry that also induces similar spatial exchange anisotropy, which supports our scenario of the local disorder in Ba$_3$Cr$_2$O$_8$. We expect that Sr$_3$Cr$_2$O$_8$ would also exhibit more than one single triplet excitation mode without field, which has very recently been confirmed by inelastic neutron scattering measurements on a single crystal of Sr$_3$Cr$_2$O$_8$\cite{Lake08}. In comparison, Ba$_3$Mn$_2$O$_8$ lacks any orbital degeneracy, and thus has spatially isotropic $J$s that leads to the single triplet excitations as observed~\cite{Stone08}.

In summary, the triplet excitations in the spin gap system, Ba$_3$Cr$_2$O$_8$, are measured by inelastic neutron scattering on single crystals while its crystal structure is determined from powder neutron diffraction experiments. The combination of these results show that Ba$_3$Cr$_2$O$_8$ is a good model system for weakly-coupled $s = 1/2$ quantum spin dimers with the intradimer coupling of $J0 = 2.38(2)$~meV and the interdimer couplings less than 0.52(2)~meV, which are spatially anisotropic due to local disorder induced by the Jahn-Teller active Cr$^{5+}$ ions.

We thank D. Khomskii, M. V. Mostovoy, C. D. Batista, S. Hass, S. Ishihara for helpful discussions.

\end{document}